\documentclass{aa}  
\usepackage{graphicx}
\usepackage{txfonts}
\usepackage{natbib}
\usepackage{aalongtable}
\newcommand{\Msun}{~M_\odot}
\newcommand{\gsim}{\raise0.3ex\hbox{$>$}\kern-0.75em{\lower0.65ex\hbox{$\sim$}}}

\begin{document}

\title{The nature of the X-Ray Flash of August 24 2005.} 
\subtitle{Photometric evidence for an on-axis 
$z=0.83$ burst with continuous energy injection and an associated supernova?\thanks{\rm This paper is based on observations from a multitude of telescopes, for example on 
observations made with ESO  Telescopes at the Paranal Observatory
(programme ID 075.D-0270) and with the NTT and ESO/Danish 1.5-m telescope at the
La Silla Observatory.
Also based on observations made with the Nordic Optical Telescope, operated
on the island of La Palma jointly by Denmark, Finland, Iceland,
Norway, and Sweden, in the Spanish Observatorio del Roque de los
Muchachos of the Instituto de Astrofisica de Canarias.
}}

\author{
J.~Sollerman\inst{1,2}
\and
J.~P.~U.~Fynbo\inst{1}
\and
J.~Gorosabel\inst{3}
\and
J.~P.~Halpern\inst{4}
\and
J.~Hjorth\inst{1}
\and
P.~Jakobsson\inst{1,5}
\and
N.~Mirabal\inst{4}
\and
D.~Watson\inst{1}
\and
D.~Xu\inst{1}
\and
A.~J.~Castro-Tirado\inst{3}
\and
C.~F\'eron\inst{1}
\and
A.~O.~Jaunsen\inst{6}
\and
M.~Jel\'{\i}nek\inst{3}
\and
B.~L.~Jensen\inst{1}
\and
D.~A.~Kann\inst{7}
\and
J.E.~Ovaldsen\inst{6}
\and
A.~Pozanenko\inst{8}
\and
M.~Stritzinger\inst{1}
\and
C.~C.~Th\"one\inst{1}
\and
A.~de Ugarte Postigo\inst{3}
\and
S.~Guziy\inst{3}
\and
M. Ibrahimov\inst{9}
\and
S.~P.~J\"arvinen\inst{10,11} 
\and
A.~Levan\inst{5}
\and
V.~Rumyantsev\inst{12} 
\and
N.~Tanvir\inst{5,13}
}
\institute{
Dark Cosmology Centre, Niels Bohr Institute, University 
of Copenhagen, Juliane Maries Vej 30, DK--2100 Copenhagen~\O, Denmark
\and
Stockholm Observatory, Department of Astronomy, AlbaNova, 
S-106 91 Stockholm, Sweden
\and
Instituto de Astrof\'\i sica de Andaluc\'\i a (IAA-CSIC), 
P.O. Box 3.004,
E-18.080 Granada, Spain
\and
Columbia Astrophysics Laboratory, 550 West 120th Street, Columbia University, New York, NY 10027-6601, USA
\and
Centre for Astrophysics Research, University of Hertfordshire, Collage Lane, Hatfield, Herts, AL10 9AB, UK
\and
Institute of Theoretical Astrophysics, PO Box 1029, N-0315 Oslo, Norway
\and
Th\"uringer Landessternwarte Tautenburg,  Sternwarte 5, D--07778 Tautenburg, Germany
\and
Space Research Institute (IKI), 84/32 Profsoyuznaya Str, Moscow 117997, Russia
\and
Ulugh Beg Astronomical Institute, Tashkent 700052, Uzbekistan
\and
Astrophysikalisches Institut Potsdam, An der Sternwarte 16, D-14482 Potsdam, Germany
\and
Astronomy Division, P.O. Box 3000, FIN-90014 University of Oulu, Finland
\and
Crimean Laboratory of Sternberg Astronomical Institute MSU, Nauchny, Crimea, 98409, Ukraine
\and
Department of Physics and Astronomy, University of Leicester, Leicester, LE1 7RH, UK
}

   \date{}

 
\begin{abstract}
{}
{Our aim is to investigate the nature of the X-Ray Flash (XRF) of August 24, 2005. 
}
{We present comprehensive photometric $R$-band observations of the fading optical afterglow of XRF\,050824,
from 11 minutes to 104 days after 
the burst. In addition we present observations 
taken during the first day in the $BRIK$ bands
and two epochs of spectroscopy. We also analyse available X-ray data.
}
{
The $R$-band lightcurve of the afterglow resembles the lightcurves
of long duration Gamma-Ray Bursts (GRBs), i.e., a power-law, albeit 
with a rather shallow slope of
$\alpha=0.6$ ($F_{\nu} \propto t^{-\alpha}$). 
Our late $R$-band images reveal the host galaxy.
The rest-frame $B$-band
luminosity is $\sim 0.5$ L$^{*}$. The star-formation rate as
determined from the [O II] emission line is $\sim1.8$~M$_{\odot}$
yr$^{-1}$.
When accounting for the host contribution, the slope is
$\alpha=0.65\pm0.01$  and a break in the lightcurve is suggested.
A potential lightcurve bump at 2 weeks can be interpreted as a 
supernova only if this is a 
supernova with a fast rise and a fast decay. 
However, the overall fit still shows excess scatter
in the lightcurve in the form of wiggles and bumps.
The flat lightcurves in the optical and X-rays could be explained by a continuous energy injection scenario, 
with an on-axis viewing angle and a wide jet opening angle ($\theta_j~\gsim~{\mbox{${10}^\circ$}}$).
If the energy injections are episodic this could potentially help explain the bumps and wiggles.

Spectroscopy of the afterglow gives
a redshift of
$z=0.828\pm0.005$ from both absorption and emission lines.
The spectral energy distribution (SED) of the afterglow has a
power-law ($F_{\nu} \propto \nu ^{-\beta}$) shape with slope
${\beta}=0.56\pm0.04$. This can be compared to the X-ray spectral index which
is ${\beta_{\rm X}}=1.0\pm0.1$. 
The curvature of the SED constrains the dust reddening towards the burst to $A_{\rm v}<0.5$ mag. 
}
{}
 \end{abstract}


\titlerunning{XRF 050824}
\maketitle

\section{INTRODUCTION}

X-Ray Flashes (XRFs) are similar to Gamma-Ray Bursts (GRBs), but with 
most of the fluence of the prompt emission
detected in the X-ray band. Their existence as a class 
was suggested by \citet{heise01} based 
on data from the BeppoSAX satellite. 

The high energy spectra ($\nu F_{\nu}$) of the prompt emission from GRBs
are well described by the 
so-called Band function \citep{band93}, which is composed of two smoothly
connected power-laws. The energy at which the two power-laws connect,
$E_{\rm peak}$, is where most of the energy 
is emitted.  
For classical GRBs, $E_{\rm peak}$ 
is typically a few 100~keV \citep{preece00}. The
spectra of the prompt emission
of XRFs are also well fitted by the Band function, but with
values of $E_{\rm peak}$ below $\lesssim50$~keV and in some cases even below 10~keV
\citep{kippen03,barraud03}. 

\citet[][]{sakamoto05} 
argue, in accordance with previous studies, that the
spectral distributions of XRFs,  X-ray Rich GRBs and GRBs form a continuum,
suggesting that they all arise from the same phenomenon \citep[see also][]{barraud03,barraud05}. 
There has also been growing evidence that (at least some) 
XRFs are the result of classical GRBs
seen off-axis \citep{yamazaki02,yamazaki03,rhoads03,fynbo04,granot05}.

Other
suggestions include XRFs as either dirty fireballs, which are relativistic
jets with a larger baryon load and hence (assuming external shocks) lower
$\Gamma$-factors than those of classical GRBs \citep{dermer99,heise01}, or XRFs
as clean fireballs where (assuming internal shocks) the spread in 
$\Gamma$-factors is small but the average $\Gamma$-factor is large
\citep{barraud05}.

However, there are still open questions regarding the origin of XRFs. 
While the connection between long GRBs and certain core-collapse 
supernovae appears to be well
established \citep{hjorth03a,matheson03,zeh04}, 
the case has not been as well
defined for XRFs.  \citet{fynbo04} performed the first comprehensive
observational campaign of an XRF optical afterglow, for XRF\,030723 \citep[see
also][]{butler05}. 
The well sampled lightcurve for XRF\,030723 displayed several interesting
features:

(i) The very early lightcurve was essentially flat, in accordance
with models for which XRFs are viewed away from the jet axis.  

(ii) The
following power-law decay was very similar to that seen in typical GRBs,
suggesting a common origin.  

(iii) At $\sim16$ days past burst a
strong bump in the lightcurve suggested the presence of a fast rising
supernova. The supernova interpretation was argued to be
consistent with the spectral energy distribution (SED) evolution
\citep{fynbo04} and was later supported by modeling both of the afterglow
\citep{granot05} and of the supernova \citep{tominaga04}.  

More evidence has now been 
presented arguing for a common origin for GRBs and XRFs. 
XRF\,020903 had a late lightcurve and spectrum consistent
with a supernova at $z=0.25$ \citep{soderberg05,bersier06}. 
More recently, the nearby \citep[$z=0.0331$,][]{mirabal06,wiersema07} 
and unusual burst, XRF\,060218, with a very low 
$E_{\rm peak}\sim5$~keV \citep{campana06} showed unambiguous evidence for an associated supernova \citep[e.g.,][]{sollerman06,modjaz06,mirabal06b,pian06}. SN\,2006aj associated with XRF\,060218 clearly established the close link between SNe, (GRBs) and XRFs.

However, other XRFs
with late time coverage did apparently not show any clear evidence for a supernova bump
\citep{soderberg05,levan05}. 
This could point to a difference in the origin of XRFs and the long GRBs with accompanying SNe \citep[but see the recent paper by][for GRBs with no associated supernova emission]{fynbo06},
although any such claim has to await observations of XRFs with a better
determined distance scale. 
Most of the XRFs claimed to lack SNe in the above-mentioned studies had no measured redshifts.

We here 
present data for XRF\,050824 for which we have obtained
a well monitored optical lightcurve 
and a secure spectroscopic redshift. 

\subsection{XRF\,050824}

XRF\,050824 was detected by the BAT instrument on-board the Swift satellite 
\citep[][]{campana05gcna} on August
24.9669 2005 UT (Universal Time is used throughout the article).
The burst duration (T$_{90}$) was 25~s, i.e., this was  a 
long-duration burst. The total 
fluence in the 15--150~keV band was $\sim$2.3$\times$10$^{-7}$ ergs cm$^{-2}$
\citep{krimm05gcn3871}.
The burst was also detected by the 
FREGATE instrument onboard HETE-2 \citep{crew05gcn}. As seen by HETE-2 
the value of the fluence ratio S($2-30$)/S($30-400$)$=2.7$. 
\citet{crew05gcn} estimated $E_{\rm peak} < 12.7$~keV.
GRB\,050824 is thus clearly an XRF.

The BAT on-board localization was reported to an accuracy of 3 arcminutes.
Early ROTSE-III data did not detect any new object \citep{schaefer05gcn}, 
but our observations
starting 38 minutes after the trigger revealed a new object at
R.A. = 00:48:56.1, Dec = +22:36:32 (J2000, see Sect.~\ref{astrometry}),
which we later confirmed as the
counterpart \citep{gorosabel05a}.

In this paper we focus on our comprehensive optical 
study of this afterglow.
The article is organized as follows. 
Sect.~\ref{observations} outlines how the optical observations were obtained and reduced. The results are presented in Sect.~\ref{results}, which includes the astrometry, the optical lightcurve, the spectral energy distribution, the spectroscopic results, data on the host galaxy and an analysis of available X-ray data of the afterglow. Finally, we end the paper with a discussion (Sect.~\ref{discussion}) 
and some conclusions (Sect.~\ref{conclusions}).

\section{OBSERVATIONS}\label{observations}

\subsection{Photometry}\label{photobservations}

Our very first observations of XRF\,050824 were obtained with the BOOTES-1B 
30~cm robotic telescope
\citep[e.g.,][]{bootes}
 in southern Spain, which detected the burst  
from 10.6 minutes after the high energy event in several $R$-band exposures.
However, since these images were not processed until 
later,
the actual discovery was instead made via the
studies we initiated 
38 minutes after the burst \citep{gorosabel05a} 
using the 1.5~meter telescope at the Observatorio de Sierra Nevada (OSN) 

We conducted a comprehensive study of the optical 
afterglow over the following 100 days using several telescopes and instruments. 
In Table~\ref{t:telescopes} we summarize the telescopes and instruments used 
and provide details on the field-of-view (FOV) and pixel scale of these instruments.

\begin{table}
\caption{Telescopes and Instruments
\label{t:telescopes}}
\centering   
\begin{tabular}{l c c l}
\hline\hline  
 Telescope & Instrument/ & FOV & Pixel scale \\

           &   CCD       & (arcminutes) & (arcsec pixel$^{-1}$) \\
\hline
BOOTES-1B & ST8E & $40\times26$ & 1.6\\
OSN & ROPER & $7.92\times7.92$ & 0.232\\
NOT & ALFOSC & $6.3\times6.3$ & 0.189 \\
NOT & STANCAM & $3\times3$ & 0.176 \\
D1.5m & DFOSC & $13.7\times13.7$ & 0.395 \\
MDM 1.3m &   SITe CCD  &   $8.6\times8.6$ &    0.508 \\
MDM 2.4m &   SITe CCD  &   $9.4\times9.4$ &    0.275 \\
CrAO2.6& FLI-IMG1001E & $8.5\times8.5$ & 0.5 \\
Maidanak1.5m & SITe CCD   &       $8.5\times3.5$ &  0.266\\
WHT & ULTRACAM & $5\times5$ & 0.3\\
NTT & EMMI & $9.9\times9.1$ & 0.33 \\
VLT & FORS1 & $6.8\times6.8$ & 0.20 \\
VLT & FORS2 & $6.8\times6.8$ & 0.25 \\
VLT & ISAAC & $2.5\times2.5$ & 0.148 \\
\hline
\end{tabular}
\end{table}

We used the ESO/Danish 1.54~m telescope (D1.5m) on La Silla 
equipped with the DFOSC instrument,
the 2.56~m Nordic Optical Telescope (NOT) on La Palma equipped with ALFOSC 
and STANCAM.
We also used the 1.3~m MDM telescope (in August 2005) and the 2.4~m MDM telescope (in September), the Crimean Astrophysical Observatory (CrAO) 2.6~m telescope and the 1.5~m telescope at the Maidanak observatory. 

Late observations were also obtained at the ESO New Technology Telescope (NTT)
at La Silla and at the Very Large Telescope (VLT) on Paranal, Chile. 
A single epoch near infrared 
(near-IR) $K$s image was obtained using the VLT/ISAAC instrument.

The full journal of observations is given in Table~\ref{t:log}.
The data were reduced using standard techniques for de-biasing and
flat-fielding.

\subsection{Spectroscopy}\label{specobservations}

Spectra of the source were obtained with the VLT at two epochs. 
 A $2\times1500$~s 
spectrum was obtained on August 25.4, about 0.4 days past the burst,
 when the afterglow had a magnitude
of R$\approx20.7$. We used the FORS2 spectrograph with a 300V grism, the GG375 order 
separation filter and a 1.0 arcsec wide slit providing a dispersion of 
13.3~\AA \ over the spectral region from 3800~\AA \ to 8900~\AA. 

The following night, on August 26.3, when the afterglow had faded by one magnitude,  we obtained another spectrum of
$6\times1500$~s exposure time. The instrumental setup was identical to that used on the first night.

We extracted the spectrum using standard procedures within {\tt IRAF}. Wavelength calibration was obtained using images of HeNeAr lamps obtained as part of the morning calibrations. Flux calibration was performed using spectra of the spectrophotometric standard star G138-31 (Oke 1990).

\section{RESULTS}\label{results}
\subsection{Astrometry}\label{astrometry}

We determined the celestial position of the XRF\,050824
optical afterglow as the mean astrometric solution found in 10 OSN
R-band  images. Each afterglow position is based on $\sim50$ USNO-
A2.0 reference stars per image. The mean value of the coordinates are:

\noindent{R.A.(J2000)= 00:48:56.14$\pm$0.03s}, \\
\noindent{Dec (J2000)=$+22$:36:33.2 $\pm 0.4^{\prime\prime}$}\\
These astrometric errors include the $0.25^{\prime\prime}$ systematic
error of the USNO-A2.0 catalogue 
\citep{assafin01,deutsch99}.

\subsection{The Lightcurve}
\begin{figure*}[h]
\begin{flushleft}
{\includegraphics[width=.5\textwidth,angle=0,clip]{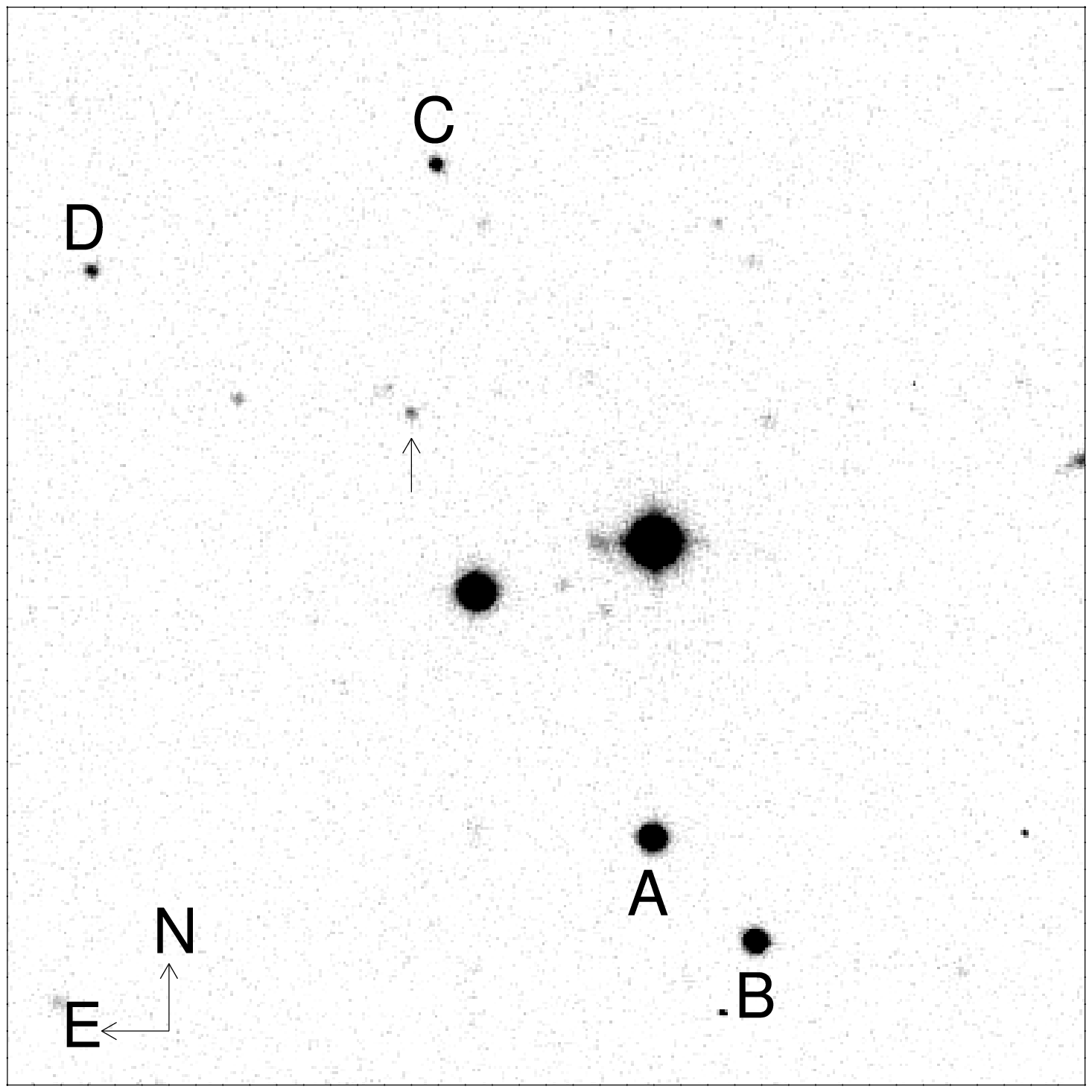}}
\caption{}{
The 156$\times$156 arcsec$^2$ field around the position of 
XRF\,050824 in a D1.5m R-band image taken 6~hr after the burst.
East is to the left and North is up.
The position of the afterglow is marked with an arrow and
the four local calibration stars are marked with capital
letters.
}
\label{f:field}
\end{flushleft}
\end{figure*}

The photometry of the XRF was carried out using 
either PSF fitting photometry (when the afterglow was bright) or
aperture photometry (at later times when the host started to contribute significantly).
We measured the magnitudes of the optical afterglow as well as 4 stars 
in the field. The relative magnitudes were transformed to the standard system 
using observations of photometric standard stars \citep{landolt}. 
The local standard stars are marked in Fig.~\ref{f:field}, and their magnitudes are given in Table~\ref{t:stars}.
The zeropoint uncertainties are of the order of 0.03 mag.

\begin{figure*}
\begin{flushleft}
{\includegraphics[width=0.5\columnwidth,angle=0,clip]{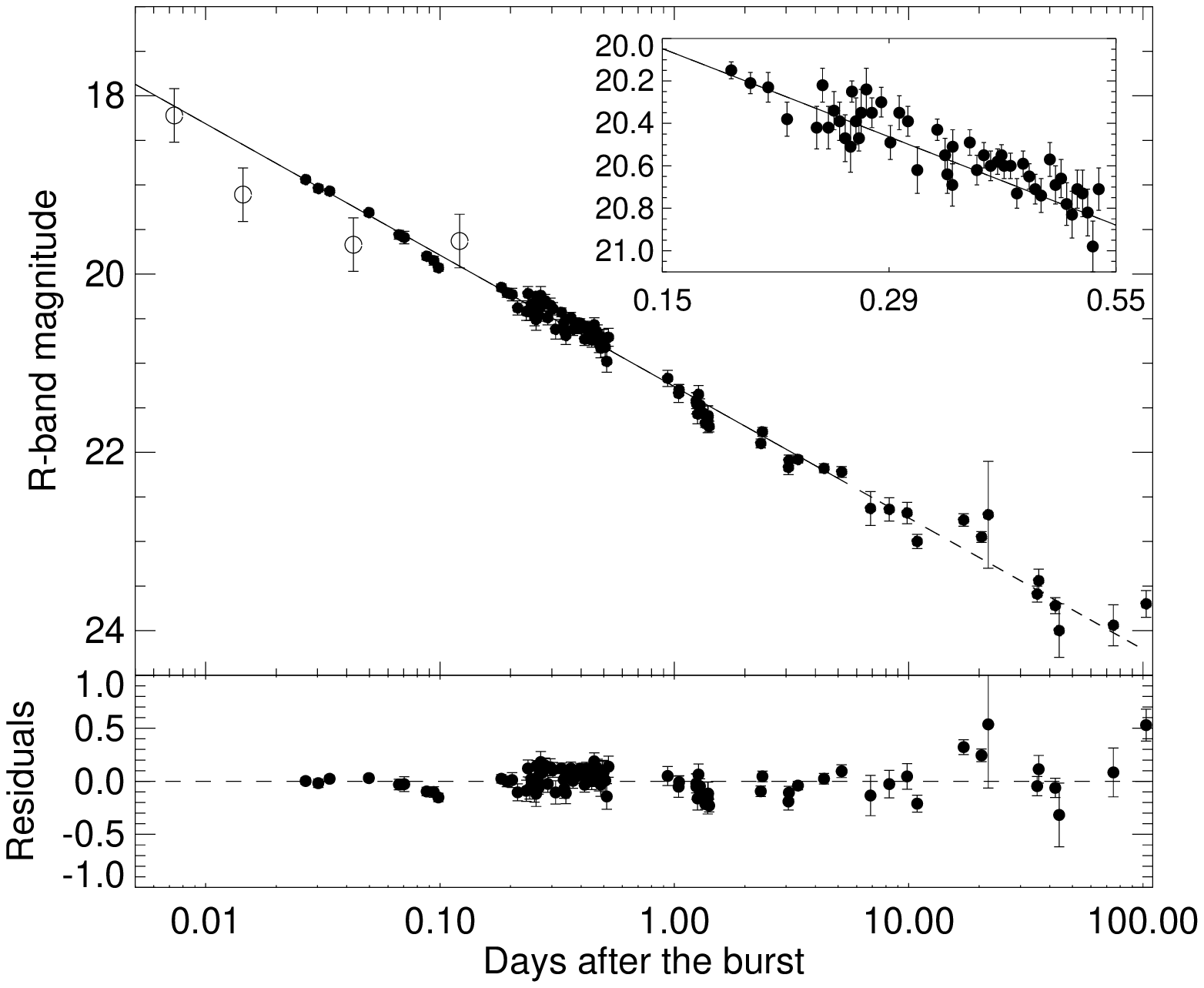}}
\caption{}{
The $R$-band lightcurve of the afterglow of XRF\,050824. 
The line represents a power-law decay with decay slope $\alpha=0.59$.
The open circles represent the BOOTES detections.
The inset highlights the well sampled phase at 0.15 -- 0.55 days, and the lower panel shows the residuals from the best power-law fit. 
}
\label{f:lc}
\end{flushleft}
\end{figure*}

In Fig.~\ref{f:lc} we have plotted the R-band 
lightcurve ranging from 11 minutes to 104~days after the XRF. 
This includes the early detections from the 
BOOTES telescope (open circles).
The best fit power-law has a slope of
$\alpha_{\rm R} = 0.59\pm0.01$ ($F_{\nu} \propto t^{-\alpha}$) 
and is also indicated in the figure.

The $I$-band lightcurve follows the $R$-band very well during the first day when we have observations in both bands. The slope is consistent with the $R$-band lightcurve ($\alpha_{\rm I} = 0.51\pm0.06$, whereas 
$\alpha_{\rm R} = 0.57\pm0.03$ for the same period).

\subsection{Spectral Energy Distribution}
\label{SED}

The multiband observations of XRF\,050824 allowed us to construct the
spectral energy distribution
(SED) of the burst at an epoch of $\sim0.4$ days.

\begin{figure*}[h]
\begin{flushleft}
{\includegraphics[width=0.5\columnwidth,angle=-90,clip]{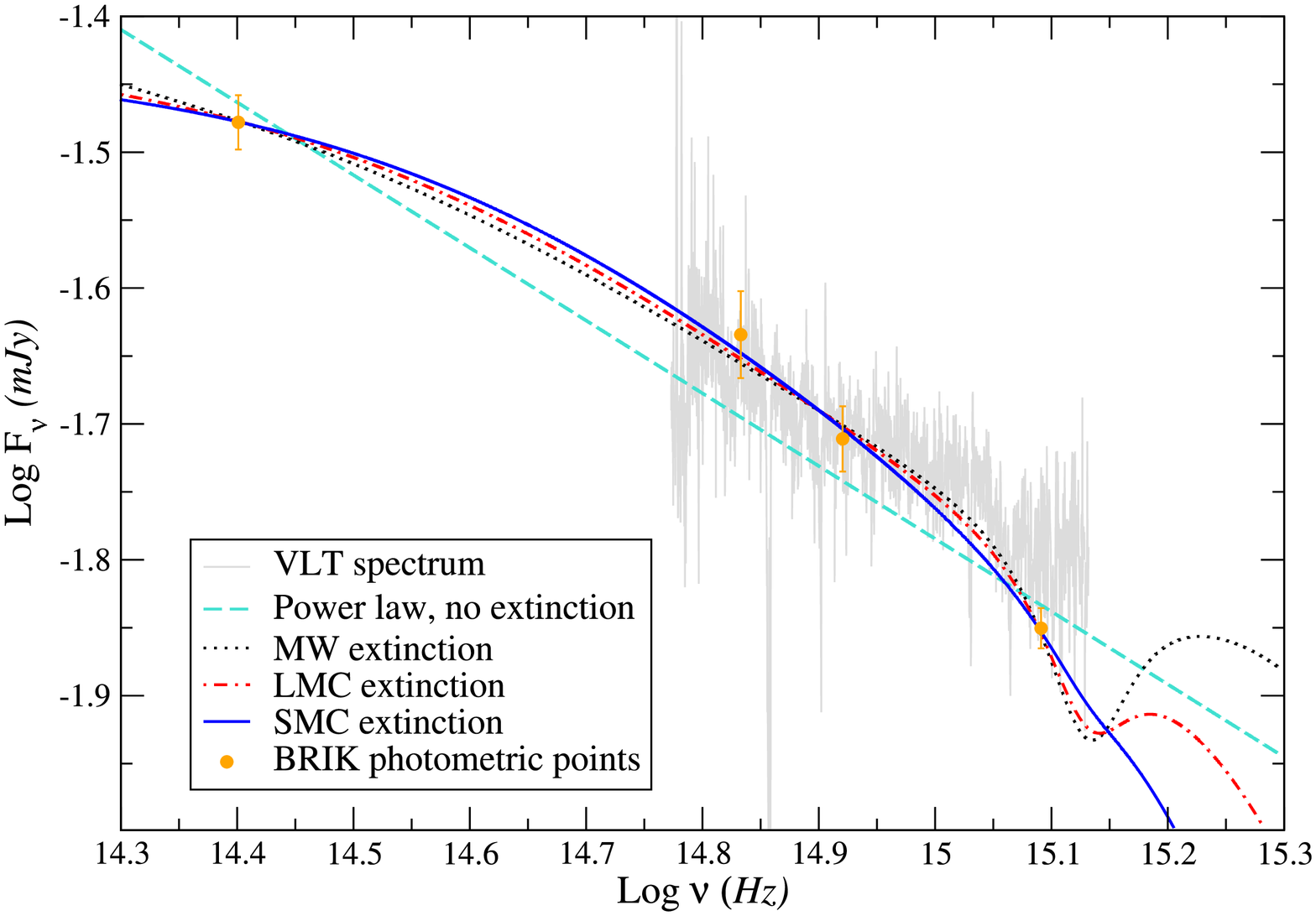}}
\caption{}{
The spectral energy distribution for XRF\,050824. 
These are the AB magnitudes in $B$, $R$, $I$,
and $K$ corrected for Galactic extinction of E$(B-V)=0.035$~mag with R$_{\rm V}$=3.1, and 
interpolated to the same epoch at $\sim0.4$ days past explosion.
}
\label{f:SED}
\end{flushleft}
\end{figure*}

The result is presented in Fig.~\ref{f:SED} where we have converted the 
$B$, $R$, $I$ and $K$ band magnitudes into AB magnitudes.
The optical and near-IR magnitudes were corrected for
Galactic reddening of E$(B-V)=0.035~$mag~\citep[][]{schlegel98} 
and transformed to flux densities using the conversion factors given
by  \citet[][]{fukugita95} and \citet[][]{allen00},
respectively. 
Given that the
multiband observations were not all performed at the same
epoch, their corresponding fluxes were rescaled using the best fit power law.

A power-law fit in the form 
F$_{\nu} \propto \nu^{-\beta}$ 
provides a tolerable fit for the SED.
The spectral index is $\beta = 0.56\pm0.04$, assuming negligible extinction.

Unfortunately, with the available data we cannot say much about the extinction.
The SMC, LMC or Milky Way extinction laws give equally good fits to the data (Fig.~\ref{f:SED}). 
Fixing the redshift at $z=0.83$ (see Sect.~\ref{spectrum}), a
free fit with an intrinsic power-law shape of the SED and an SMC like extinction curve from Pei (1992) implies an extinction of
$A_{\rm V}=0.4\pm0.2$~mag. However, this would give an unrealistically flat $\beta\sim0$ spectrum.
Given the sparse dataset the only thing we can firmly conclude is that A$_{\rm V}$ is less than 0.5 mag. A low value of the global extinction is also implied by the blue color of the host galaxy (Sect.~\ref{host}).

\subsection{The Spectra}\label{spectrum}

The combined flux-calibrated spectrum is also included in Fig.~\ref{f:SED}.
The slope is consistent with the contemporary photometry.

\begin{figure*}[h]
\begin{flushleft}
{\includegraphics[width=0.50\columnwidth,angle=0,clip]{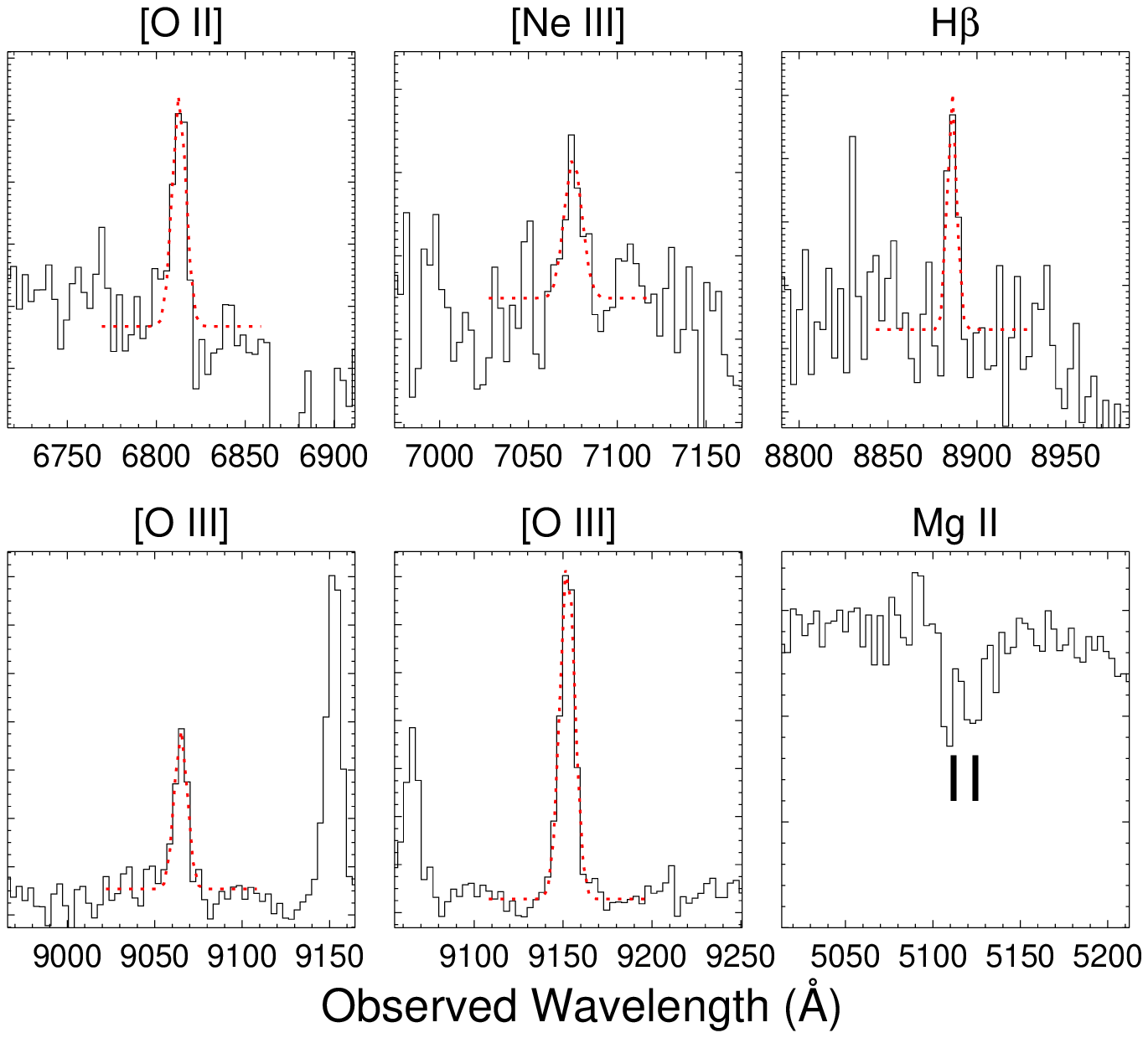}}
\caption{}{Emission lines from the VLT spectra and overplotted Gaussian fits. We refer to Table~\ref{t:lines} for the measurements. Lower right panel shows the Mg II absorption lines. The tickmarks indicate the expected positions for this doublet line given the redshift determined from the emission lines.
}
\label{f:lines}
\end{flushleft}
\end{figure*}

We determined the redshift
from the first night's spectrum 
\citep{fynbo05gcn} using emission lines such as 
[\ion{O}{II}] $\lambda 3727$, 
[\ion{O}{III}] $\lambda\lambda 4959, 5007$ 
and H$\beta$. As noted by \cite{fynbo05gcn}, we also detect 
absorption lines from \ion{Mg}{II} at this redshift. 
We discuss this further below (Sect.~\ref{abslines}).
The lines are shown in Fig.~\ref{f:lines} and the measured 
positions and fluxes of the lines are given in Table~\ref{t:lines}.
The redshift is $z=0.828\pm0.005$. Note that the fluxes given in Table~\ref{t:lines} are not corrected for Galactic or host galaxy extinction. The uncertainties in absolute line fluxes can be up to 30$\%$.  

Using a  
cosmology where $H_0=70$\,km\,s$^{-1}$\,Mpc$^{-1}$, $\Omega_\Lambda = 0.7$
and $\Omega_{\rm m}=0.3$,
this redshift corresponds to a luminosity distance of 5.24 Gpc. This distance will be used hereafter.

\subsection{The Host Galaxy}\label{host}

\begin{figure*}[h]
\begin{flushleft}
{\includegraphics[width=0.5\columnwidth,angle=0,clip]{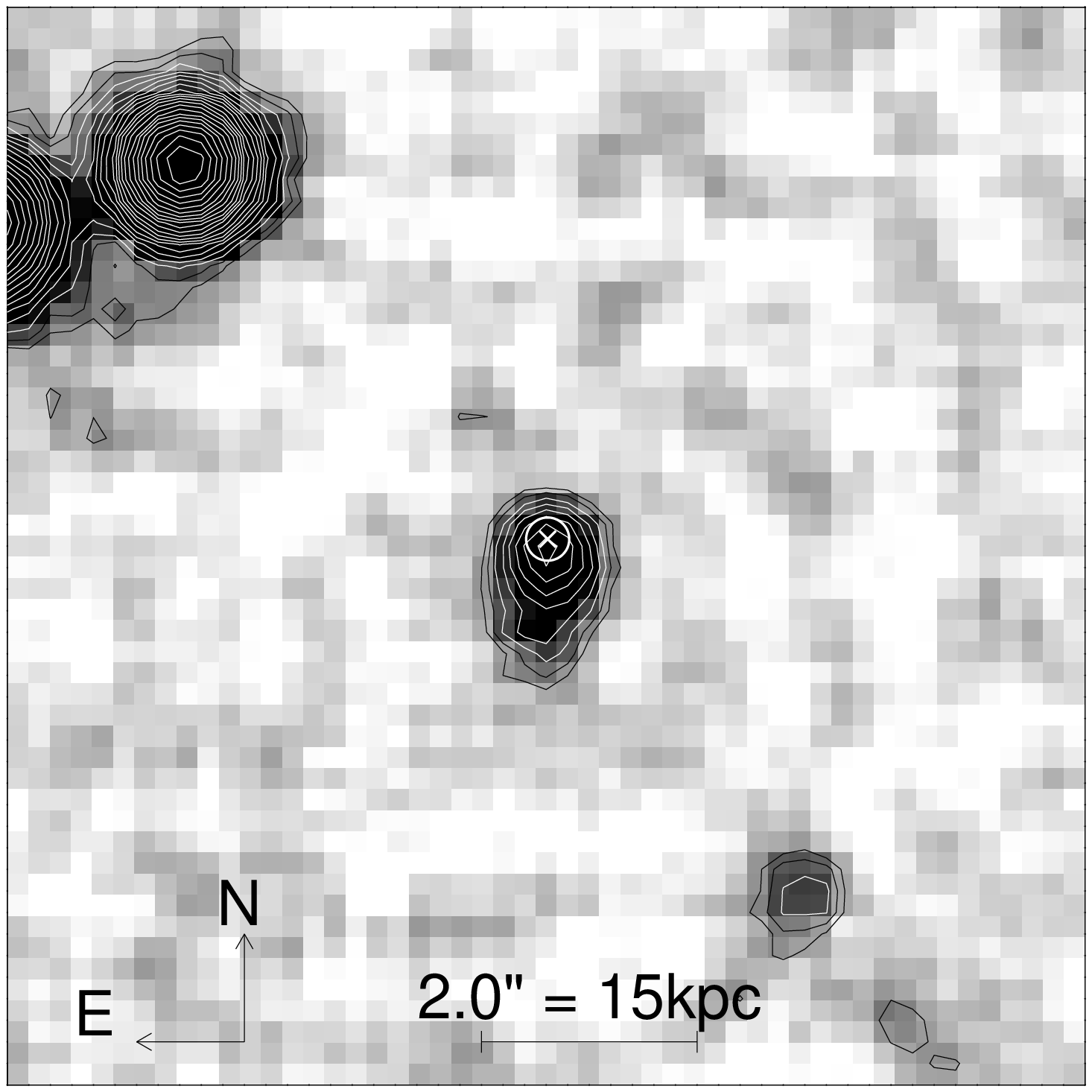}}
\caption{}{
The 10$\times$10 arcsec$^2$ field around the position of 
XRF\,050824 from our latest VLT $R$-band image 42 days after the burst.
East is to the left and North is up.
The position of the afterglow is marked with a cross and the 
3-sigma error circle on the position of the afterglow.
The host has a magnitude of $R$=23.7. 
The FWHM of pointlike objects is 0.5 arcsec in this image.
}
\label{f:host}
\end{flushleft}
\end{figure*}

Our late $R$-band image from December 7, 2005, 104 days past explosion, shows
an extended source with magnitude $\mathrm{R}=23.70\pm0.15$ 
at the position of the afterglow. This is the host galaxy of XRF\,050824.
An image obtained under very good seeing conditions at VLT in October 2005 is shown in Fig.~\ref{f:host}. Note that all late observations (past 35 days) are consistent with this being the host, with little contribution from the afterglow (or a supernova).

The host magnitude is 23.6 when corrected for a Galactic extinction of $E(B-V)=0.035$~mag. 
At the redshift of this galaxy
the $R$ band corresponds to the rest frame $U$ band.
However, comparison of the absolute luminosity with other galaxies is often made in the rest frame $B$ band. To do so we
need to make some assumption about the color of the host. Here we note that the $BVR$ magnitudes from our latest VLT data, when the host is clearly dominating the emission, are very similar to the magnitudes of the host of GRB\,000210 \citep{gorosabel03} at a similar redshift. 

We therefore conclude that the absolute luminosity of the XRF\,050824 host is very similar to the one determined by 
\citet{gorosabel03}, i.e., $L=0.5\pm0.2~L_{\star}$ in the rest frame $B$ band. 

The host is extended with a size of roughly 0.8 arcsec,
which at a distance of 
5.24 Gpc corresponds to a linear scale of $\sim6$~kpc. 
Finally, from the [O~II] lines we can estimate the star formation rate (SFR). 
The flux of this line, corrected for Galactic extinction, 
corresponds to a SFR of $1.8~\Msun$~yr$^{-1}$, following \cite{kennicutt98}.

In fact, using the extinction corrected value for the flux of H$\beta$, and assuming a case B recombination ratio for H$\alpha$ versus H$\beta$, we can also use this line to estimate that the SFR is $1.8~\Msun$~yr$^{-1}$, again following \cite{kennicutt98}. As usual, any slit-losses would increase this number.  We note that the consistency of the H$\alpha$ and [O~II] predictions of the 
SFR also supports the notion of low extinction in the host galaxy.

This SFR compares rather well with the estimate for the host of GRB\,000210 mentioned above, which has similar properties and an estimated SFR from the UV light of $~2.1\pm0.2~\Msun$~yr$^{-1}$ \citep{gorosabel03}.

The specific star formation rate for the host galaxy of XRF\,050824 
is thus only 
$\sim4~M_{\odot}$ yr$^{-1} (L/L^{\star})^{-1}$.
This is rather low, but not exceptional, and falls well within the population
of small star-forming blue galaxies as shown in Fig.~2 of \citet{sollerman05}, 
\citep[see also][]{christensen04}.

Finally, we can estimate the metallicity of the galaxy using the R23 technique
\citep{pagel79}. Using the results presented in Table~\ref{t:lines} and
applying E$(B-V)=0.035$~mag, we derive log(R$_{23}$)=1.0. This is quite high,
and indicates a metallicity (just) below solar \citep[see e.g., Fig.~5
by][]{kewley02}. However, the small number of emission lines in the analysis 
makes this estimate rather uncertain.

The star formation rate and size thus indicates a fairly normal galaxy,
similar to other GRB host galaxies
\citep[e.g.,][]{lefloch03,christensen04,sollerman05}. The metallicity 
confirms the trend that GRB host galaxies have sub-solar metallicities. The
luminosity is not particularly low compared to other GRB hosts, but is similar to the host of XRF\,050416A 
\citep{soderberg06}.

\subsection{The X-rays}

Swift-XRT did not observe the burst immediately due to a lunar
constraint and the XRT began observations about 6000\,s after the
burst trigger \citep{campana05gcnb}. We have analysed the standard processed 
XRT data starting at 
0.4 days after the burst 
using version 2.3 of the Swift software. Background-subtracted spectra
and lightcurves were extracted in a standard way with circular source
and background extraction apertures of 30\arcsec\ and 60\arcsec\
radius for the PC-mode data. Data from the WT-mode were not used because
they added very little signal.

The combined spectrum was fit with a single absorbed power-law with
absorption at the Galactic level \citep[$N_{\rm H} =
3.62\times10^{20}$\,cm$^{-2}$,~][]{dickey90} and gave an
acceptable fit ($\chi^2=43.7$ for 40 degrees of freedom). 
Adding an absorber at the redshift of the host galaxy
gives a better fit
($\chi^2=30.4$ for 39 degrees of freedom) with an equivalent
hydrogen column density of $N_{\rm H} =
1.8^{+0.7}_{-0.6}\times10^{21}$\,cm$^{-2}$ and a power-law photon
index $\beta_{\rm X}=1.0\pm0.1$.
The absorption
model has abundance ratios fixed at solar values. The soft X-ray
absorption is dominated by $\alpha$-chain elements and is therefore a
measure of the metal absorption and is regardless of whether the
elements are in the gas or solid phase 
\citep[see][]{watson06,turnshek03}.

The flux decay of the afterglow followed a single power-law with decay
index, $\alpha_{\rm X}=0.75\pm0.04$, 
with the fit
being marginally acceptable ($\chi^2=24.8$ for 16 degrees of freedom,
null hypothesis probability = 0.07). This decay rate is somewhat faster than the average slope seen in the optical lightcurve (albeit there could be a break in the optical light curve, see Sect.~\ref{break}). 
Note, however, that the X-ray data are not very constraining at the later phases.

\section{DISCUSSION}\label{discussion}

\subsection{Absorption line redshift}
\label{abslines}

As noted in Sect.~\ref{spectrum} the spectra also include absorption lines from Mg II. These lines are seen at both epochs, but are most clearly detected in the first epoch, which has the best signal-to-noise ratio. The lines are displayed in Fig.~\ref{f:lines}, and the redshift is consistent with the estimate from the emission lines.

That the redshift can be determined from both emission lines and absorption lines is of some importance. The distance scale of the XRFs has only recently been 
shown to be cosmological. The first spectroscopic redshift of $z=0.25$ 
for  XRF\,020903 \citep{soderberg05} was based on emission lines only. In principle, a single case could be affected by a superposed and unrelated galaxy, but now  XRF\,050416A also has a measured emission line redshift \citep[$z=0.65$,][]{cenko05,soderberg06} as has GRB/XRF\,030528 \citep[$z=0.78$,][]{rau05}.

The GRB\,030429 studied by \cite{jakobsson04} also displayed absorption lines. It showed an $E_{\rm peak}$ of 35~keV at a redshift of $z=2.66$, and is thus a borderline case, consistent with an X-ray Rich burst. More recently, the rather unusual XRF\,060218 had a secure redshift from both emission lines and absorption lines \citep[$z=0.033$,][]{mirabal06,wiersema07,pian06}. 

These findings, together with the robust redshift determination for the rather normal XRF\,050824, have therefore proven the cosmological distance scale for these objects beyond doubt.

\subsection{The Lightcurve of the Afterglow}

Our $R$-band lightcurve of XRF\,050824 is one of the best sampled optical lightcurves for an XRF. The most conspicuous aspect of this lightcurve is that it is basically consistent with a power-law for the entire duration (Fig.~\ref{f:lc}). The best fit power law $\alpha = 0.6$ is quite a slow decay.

\subsubsection{The Early Times}

\begin{figure*}[h]
\begin{flushleft}
{\includegraphics[width=0.7\columnwidth,angle=0,clip]{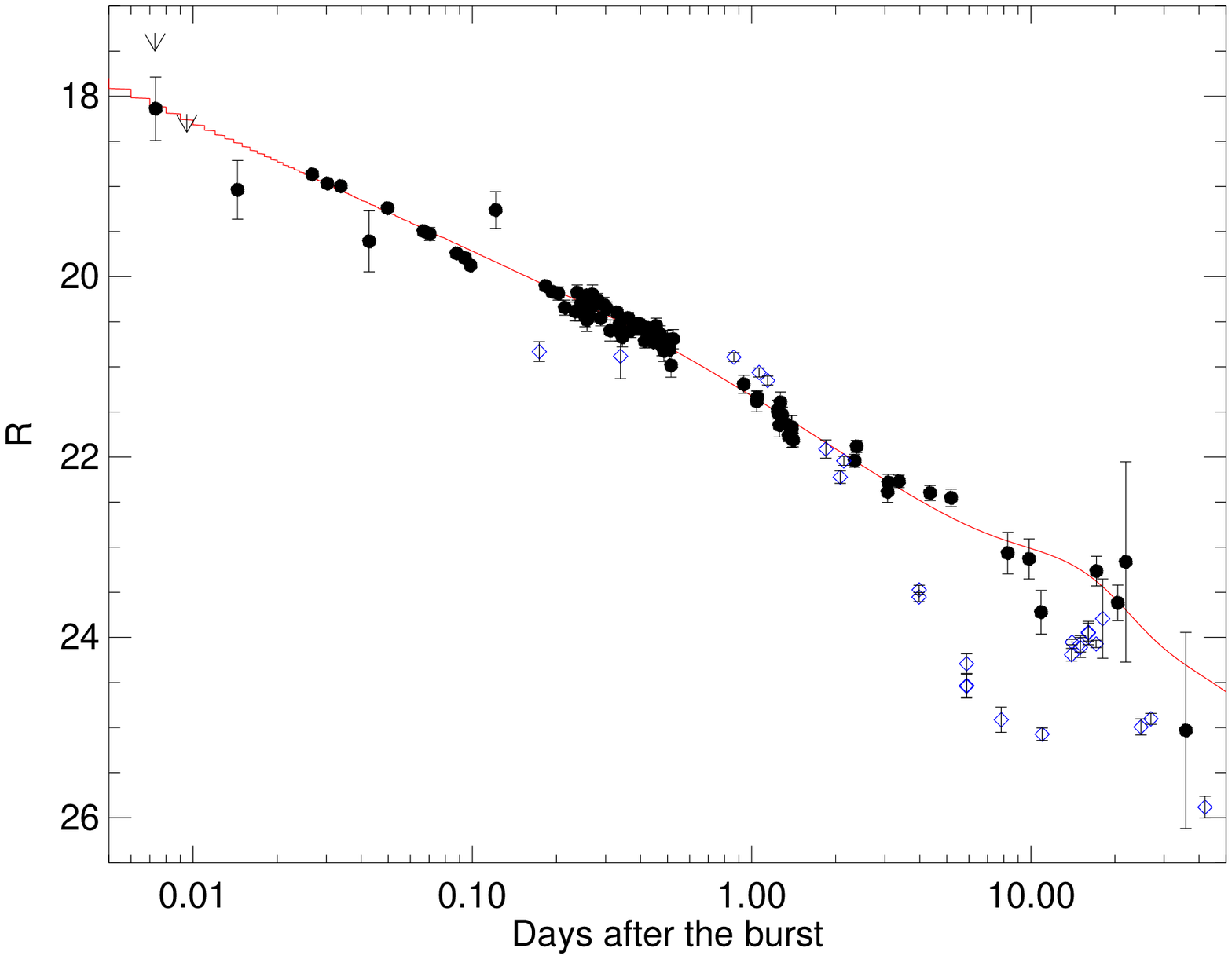}}
\caption{}{The R-band magnitudes for XRF\,050824 corrected for Galactic extinction and for the contribution from the host galaxy (black, filled dots). The (red) line is our fit to these data, following \citet{zeh04}. There is a shallow break, from $\alpha$=0.6 to 0.8 at $\sim0.5$ days. The best fit supernova has k$=1.05\pm0.42$ and s$=0.52\pm0.14$ - a bright and fast lightcurve.
The (blue) open dots are the observations of XRF\,030723 \citep{fynbo04}. These are just as observed, i.e., no assumption has been made about the redshift of that burst, although we have corrected for Galactic extinction of $E(B-V)=0.03$ mag. We note the flatter early lightcurve for XRF\,030723, the steeper late decay and the conspicuous supernova bump at 20 days past burst.
}
\label{f:sn}
\end{flushleft}
\end{figure*}

One of the more interesting aspects of the lightcurve is that it declines
steadily from very early on. This is in stark contrast to the lightcurve of
XRF\,030723 (Fig.~\ref{f:sn}), which apparently had a constant lightcurve for the
first day after the burst. 

The flat early part of the lightcurve of XRF\,030723 was interpreted in terms
of geometry \citep{fynbo04}, where an off-axis orientation can make an
increasingly large fraction of the jet visible and thus maintain a constant
(or even brightening) lightcurve \citep{granot05}. We see no evidence for this
in XRF\,050824. It's early lightcurve is consistent with the same decay seen
throughout the lightcurve. This can thus be seen as evidence for an on-axis
burst, which would mean that a geometrical interpretation does not explain the
difference between XRFs and GRBs in all cases \citep[see also][]{soderberg06}.

\subsubsection{A Break and a Bump in the Lightcurve}
\label{break}

At first glance the $R$-band lightcurve shown in Fig.~\ref{f:lc} appears consistent with a single power-law decline throughout the entire afterglow. However, since the final points are due to the host galaxy, the data do suggest a break in the lightcurve. As mentioned in Sect.~\ref{host} the host is extended, so most of the light at these epochs is indeed from the galaxy. This means that the single power law must be broken 
at an early time,
or the later points would have been much brighter. The break in the lightcurve also means that an extra component is needed to explain the excess light at $10-20$ days past the burst.

We embarked on simultaneously fitting two power-laws and a stretchable SN\,1998bw template, in accordance with the method outlined by \citet{zeh04}. Fixing the host galaxy magnitude, we were able to constrain a shallow break in the lightcurve to occur at $\sim0.5$ days past burst (Fig.~\ref{f:sn}). This could be a cooling break. The required supernova is rather unusual. In the notation of \citet{zeh04} this is a supernova with k$=1.05\pm0.42$ and s$=0.52\pm0.14$. 
This is a bright and fast lightcurve and is different from the lightcurve of the canonical 
SN\,1998bw, which is 
often associated with long GRBs, but is 
similar to, although brighter than, the
supernova associated with XRF\,060218.

The actual peak luminosity of the potential supernova is, however, 
highly uncertain.
If there is internal extinction in the host galaxy the corresponding supernova would have to be brighter, but our SED analysis shows that this can not be a very large effect. This is also supported by the rather blue color of the host galaxy, and by the deduced balmer line ratios. A larger uncertainty  arises from the assumptions on the contribution from the afterglow. With SN\,1998bw ejecting $\sim0.5~\Msun$ of $^{56}$Ni \citep[e.g.,][]{woosley99,sollerman00}, we can estimate that a supernova associated with XRF\,050824
would have had to eject at least on the order of $0.6\pm0.3\Msun$ of $^{56}$Ni. This is assuming that the peak brightness scales with nickel-mass.

\subsubsection{Wiggles, bumps, humps and jitter}
\label{scatter}

In fact, the fit to the lightcurve is not very satisfactory even after invoking a break and a hypothetical supernova bump. This is seen in Fig.~\ref{f:sn}, where the reduced $\chi^{2}$ is 1.8. The entire lightcurve of XRF\,050824 displays wiggles and humps throughout the time of the observations.  The largest deviations are from a systematic dip in the NOT data relative to the overall fit just before
0.1 days, and an increase in the scatter from 2 -- 5 days.  
These deviations cannot be
satisfactorily fitted by a broken power-law scheme that is supposed to 
model a simple impulsive shock or jet.

Another example is at 0.2 -- 0.5 days past the burst when we have a well monitored lightcurve, in particular from the  MDM. 
This is shown in the inset in Fig.~\ref{f:lc}.
A linear decay is not a formally good fit to these data. There appear to be wiggles around the steady linear decline. 
Since most of these observations were obtained at the same telescope, and have been reduced in the same way against the same local standards, we do not believe this is purely an instrumental effect. Although the statistical significance is rather low in our lightcurve, we note that similar jittering has been observed previously in GRBs, both in long GRBs \citep[e.g.,][]{gorosabel06,matheson03} and in short GRBs \citep{hjorth07}, and can be interpreted in terms of variations in the surrounding circumburst medium or as due to prolonged activity of the central engine. Similar explanations can thus be put forward also for this XRF.

\subsubsection{Continuous energy injection}
\label{energyinjection}

\begin{figure*}
\includegraphics[angle=0,width=0.5\columnwidth,clip=]{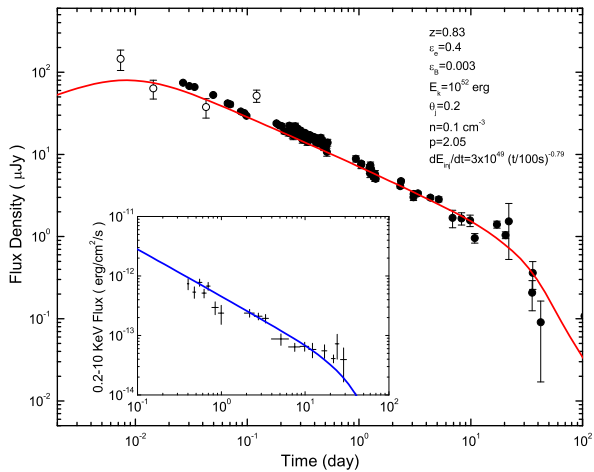}
\caption{Modeling $R$-band (host subtracted) 
and X-ray (inset) afterglow lightcurves of 
XRF\,050824 with continuous energy injection. 
The fitting parameters are given in the figure and are further
discussed in the text. 
}
\label{f:xu}
\end{figure*}

Prolonged central engine activity with multiple energy injections could thus be the explanation for the deviations from a perfect power-law \citep{bjornsson04}.
Prolonged activity in terms of continuous energy injection could also explain the slow decline rates in both optical ($\alpha_{\rm R}=0.65$ when corrected for host galaxy contribution) and in X-rays ($\alpha_{\rm X}=0.75$).
In Fig.~\ref{f:xu} we show an example of a model for the afterglow emission in the X-ray and the $R$ band for continuous energy injection. This model is detailed below.

We have modeled the afterglow in terms of 
a long-term continual energy injection in the forward shock. 
We consider an uniform relativistic jet undergoing the energy injection 
from the central source and sweeping up its surrounding uniform medium. 
The 
dynamical evolution of the outflow is calculated using the formulae in 
\citet{huang00}
and adding an energy injection process with the form  
$dE_{\rm
inj}/dt =A (t/t_0)^{-q}~{\rm for}~t_0<t<t_{\rm end}$. 
The fractions of shock energy given to the electrons $\epsilon_e$ and to the 
magnetic field $\epsilon_B$ are assumed to be constant.

The model fits shown in Fig.~\ref{f:xu}
have the 
following jet parameters: the isotropic kinetic energy $E_k=10^{52}\,{\rm erg}$,
$\epsilon_e=0.4$, $\epsilon_B=0.003$, the circumburst density $n=0.1\,{\rm
cm^{-3}}$, the electron index $p=2.05$, the half-opening angle $\theta_j=0.2$, and the viewing angle 
$\theta_{\rm
obs}$=0 (i.e., on-beam viewing), together with the energy injection 
parameters:
$A=3\times 10^{49}\,{\rm erg/s}$, $q=0.8$, $t_0=100$ s, and $t_{\rm 
end}=2\times
10^6$ s. 
This rather large amount of ejected energy is needed to explain the long and shallow decline; the amount is similar to that found for GRB\,050315 \citep{zhang06}.
Since there is no proper jet break until possibly after a week, the constraint on the jet opening angle of $\theta_j~\gsim~{\mbox{${10}^\circ$}}$
 is quite robust.
We did not attempt to fit the very early lightcurve. At these phases it is likely that a reverse shock component is required.

We note again the late re-brightening.
At such a late time, the ejecta is only moderately relativistic. 
The patchy jet model may be unable
to account for these variabilities 
\citep{kumarpiran00}, 
which may instead
be attributed to the re-activity of the central engine 
\citep[e.g.,][]{fan05},
or, as mentioned above, to a supernova (Sect.~\ref{break}).

\subsection{Amati relation}
\label{amati}

Several XRFs have been shown to follow the same relation as GRBs, that
$E_{\rm peak} \propto E_{\rm iso}^{1/2}$ \citep{amati02}, where
$E_{\rm iso}$ is the isotropic-equivalent radiated energy. 
For XRF\,050824, 
we can not determine a precise total energy due to the
lack of knowledge concerning the peak energy
of the BAT spectrum. 
However, we can calculate lower and upper limits by integrating
the best fit power law spectral energy distributions in the $(15 - 150) \times (1+z)$ keV band and in the the full $1-10^4$ keV band. 
We obtained $4.1 \times 10^{50}\,{\rm erg} < E_{\rm iso} < 3.4 \times
10^{51}\,{\rm erg}$, which when using the Amati relation, would
provide a constraint on the observed peak energy $11 \,{\rm keV} <
E^{\rm obs}_{\rm peak}< 32 \,{\rm keV}$. This 
is thus only just in agreement with that from the spectral fitting,
$E^{\rm obs}_{\rm peak}<13 \,{\rm keV}$. 

Besides this event, the Amati relation is also applicable to XRFs\,020903
and possibly 030723 \citep{lamb05},
XRF\,050416A
\citep{sakamoto05}
and XRF\,050406 \citep{schady06}. 
\citep[See also][for XRF\,060218.]{amati06,ghisellini06}

That this relation holds for both XRFs and classical GRBs not only
implies that both classes of bursts can be on-axis events 
\citep{amati06}
but also supports the idea that both can be interpreted under a unified physical
mechanism.

\section{CONCLUSIONS}\label{conclusions}

We have seen that the afterglows of XRFs can appear quite different. The early flat optical lightcurve of XRF\,030723 was consistent with predictions of an off-axis burst \citep{fynbo04,granot05}. On the contrary, XRF\,050824 displays an optical lightcurve which is decaying at a fairly constant, but slow, rate from 10 minutes after the burst. Our afterglow model indicates this to be an on-axis burst.
We have also found some evidence for a bump in the lightcurve, which is consistent with a supernova as fast as that associated with XRF\,060218, i.e, considerably faster than the SN\,1998bw lightcurve.

Most well observed XRFs with a redshift where a supernova could be expected to emerge do show some evidence for this. This is similar to the case for ordinary long GRBs \citep{zeh04}. A common origin for XRFs and GRBs is therefore likely but there also seems to be several parameters affecting the observables of the burst. In the context of the four-field diagram presented by \cite{sollerman06}, XRF\,050824 should be in the same category as XRF\,020903 and XRF\,030723; XRFs with an associated supernova but where the optical light is dominated by the afterglow at early phases. 

The mounting evidence for supernovae in GRBs and XRFs also shows that there is a large variety in supernova properties \citep{woosleybloom06}. The emergence of supernovae much fainter \citep{pian06,sollerman06} and much faster (this work) than the canonical SN\,1998bw put constraints on the underlying explosion model. Recently, \citet{fynbo06} also reported two GRBs where no supernova emission is seen, down to 100 times fainter than SN\,1998bw.
 It is still not clear whether we see two (or more) fundamentally different explosion mechanisms, or if there is a very wide continuum of properties for these blasts.

\begin{acknowledgements}
This paper is based on observations collected by the Gamma-Ray Burst 
Collaboration at ESO (GRACE) at the European Southern Observatory, 
Paranal and La Silla, Chile. We thank the ESO staff for their help in 
securing the service mode data reported here. 
We acknowledge benefits 
from collaboration within the EU FP5 Research Training Network 
"Gamma-Ray Bursts: An Enigma and a Tool". This work was also 
made at the DARK Cosmology Centre funded by The
Danish National Research Foundation.
JS acknowledges support from Danmarks Nationalbank, from the Anna-Greta and 
Holger Crafoord fund and from Knut \& Alice Wallenberg foundation.
CrAO and IKI acknowledge support from Russian Ministry of Education and
Science.
JPH and NM acknowledge support from the National Science 
Foundation under grant 0206051.
The research activities of JG and MJ are supported by the Spanish  
Ministry of Science through projects AYA2004-01515 and ESP2005-07714-C03-03.
Based in part on observations made with the BOOTES instruments in South Spain (ESAt-INTA/CEDEA, Huelva)  and  with the 1.5m  Telescope of Observatorio de Sierra Nevada (OSN), operated by IAA/CSIC.
Some of the data presented here have been taken using ALFOSC, which is owned by the Instituto de Astrof\'\i sica de Andaluc\'\i a (IAA) and operated at the Nordic Optical Telescope under agreement between IAA and the Depertment of Astronomy of Copenhagen University.
Thanks to Tamara Davis for very careful reading of the manuscript, and for detailed discussions on K-corrections.
Last but not least, we acknowledge support from several key observers that contributed to this work, namely
V. Birykov and D. Sharapov 
as well as 
D.~R.~Rafferty \&  J.~R.~Thorstensen.
Finally the late Hugo E. Schwarz contributed with observations and comments on an earlier version of this paper. 
\end{acknowledgements}

\begin{longtable}{lccccc}
\caption{\label{t:log} Log of observations and photometry of the afterglow of XRF 050824.} \\
\hline\hline
Date &  $\Delta$t & Magnitude &
Magnitude Error & Pass Band & Telescope \\
(UT) & (days) &  & ($1\sigma)$ &  & \\  
\hline
\endfirsthead
\caption{continued.}\\
\hline\hline
Date &  $\Delta$t & Magnitude &
Magnitude Error & Pass Band & Telescope \\
(UT) & (days) & & ($1\sigma)$ & & \\ 
\hline
\endhead
\hline
\endfoot
\hline
Aug   25.0973    &  0.130449  &   20.72   &  0.04 & B-band & NOT \\
Aug   25.1017    &  0.134850  &   20.72   &  0.05 & B-band & NOT \\
Aug   25.1061    &  0.139250  &   20.65   &  0.04 & B-band & NOT \\
Oct   6.1943    &  42.2274   &   24.43   &  0.16 & B-band & VLT \\
\hline
Oct   6.1898    &  42.2229   &   24.26   &  0.17 & V-band & VLT \\
\hline
Aug   24.9742    &  0.007346  &   18.22   &  0.35 & R-band & BOOTES \\
Aug   24.9813    &  0.014429  &   19.11   &  0.32 & R-band & BOOTES \\
Aug   24.9935    &  0.0266991 &   18.94   &  0.03 & R-band & OSN \\
Aug   24.9971    &  0.0302391 &   19.04   &  0.04 & R-band & OSN \\
Aug   25.0007    &  0.0338001 &   19.07   &  0.03 & R-band & OSN \\
Aug   25.0096    &  0.042705  &   19.67   &  0.33 & R-band & BOOTES \\
Aug   25.0166    &  0.0496998 &   19.31   &  0.03 & R-band & OSN \\
Aug   25.0337    &  0.0668297 &   19.56   &  0.05 & R-band & OSN \\
Aug   25.0372    &  0.0703697 &   19.59   &  0.07 & R-band & OSN \\
Aug   25.0546    &  0.0877495 &   19.80   &  0.04 & R-band & NOT \\
Aug   25.0609    &  0.0940495 &   19.85   &  0.05 & R-band & NOT \\
Aug   25.0653    &  0.0984497 &   19.93   &  0.04 & R-band & NOT \\
Aug   25.0881    &  0.121212  &   19.33   &  0.20 & R-band & BOOTES \\
Aug   25.1496    &  0.182749  &   20.15   &  0.04 & R-band & OSN \\
Aug   25.1599    &  0.193050  &   20.21   &  0.05 & R-band & OSN \\
Aug   25.1700    &  0.203150  &   20.23   &  0.07 & R-band & OSN \\
Aug   25.1813    &  0.214449  &   20.38   &  0.08 & R-band & OSN \\
Aug   25.2003    &  0.233429  &   20.42   &  0.10 & R-band & MDM \\
Aug   25.2042    &  0.237391  &   20.22   &  0.08 & R-band & MDM \\
Aug   25.2082    &  0.241360  &   20.42   &  0.10 & R-band & MDM \\
Aug   25.2122    &  0.245300  &   20.34   &  0.09 & R-band & MDM \\
Aug   25.2161    &  0.249279  &   20.39   &  0.09 & R-band & MDM \\
Aug   25.2201    &  0.253250  &   20.47   &  0.11 & R-band & MDM \\
Aug   25.2241    &  0.257210  &   20.51   &  0.12 & R-band & MDM \\
Aug   25.2251    &  0.258249  &   20.25   &  0.05 & R-band & D1.5m \\
Aug   25.2280    &  0.261179  &   20.39   &  0.11 & R-band & MDM \\
Aug   25.2303    &  0.263451  &   20.47   &  0.06 & R-band & D1.5m \\
Aug   25.2320    &  0.265150  &   20.35   &  0.11 & R-band & MDM \\
Aug   25.2360    &  0.269110  &   20.24   &  0.10 & R-band & MDM \\
Aug   25.2402    &  0.273399  &   20.35   &  0.07 & R-band & MDM \\
Aug   25.2477    &  0.280849  &   20.30   &  0.07 & R-band & MDM \\
Aug   25.2551    &  0.288280  &   20.49   &  0.08 & R-band & MDM \\
Aug   25.2626    &  0.295719  &   20.35   &  0.08 & R-band & MDM \\
Aug   25.2700    &  0.303150  &   20.39   &  0.07 & R-band & MDM \\
Aug   25.2783    &  0.311409  &   20.62   &  0.11 & R-band & MDM \\
Aug   25.2966    &  0.329741  &   20.43   &  0.05 & R-band & MDM \\
Aug   25.3040    &  0.337179  &   20.55   &  0.08 & R-band & MDM \\
Aug   25.3062    &  0.339350  &   20.64   &  0.09 & R-band & D1.5m \\
Aug   25.3111    &  0.344250  &   20.69   &  0.10 & R-band & D1.5m \\
Aug   25.3115    &  0.344610  &   20.51   &  0.08 & R-band & MDM \\
Aug   25.3288    &  0.361910  &   20.49   &  0.06 & R-band & MDM \\
Aug   25.3362    &  0.369339  &   20.62   &  0.07 & R-band & MDM \\
Aug   25.3436    &  0.376780  &   20.55   &  0.06 & R-band & MDM \\
Aug   25.3511    &  0.384220  &   20.60   &  0.07 & R-band & MDM \\
Aug   25.3587    &  0.391870  &   20.58   &  0.06 & R-band & MDM \\
Aug   25.3632    &  0.396349  &   20.55   &  0.05 & R-band & VLT \\
Aug   25.3661    &  0.399300  &   20.60   &  0.06 & R-band & MDM \\
Aug   25.3736    &  0.406740  &   20.60   &  0.06 & R-band & MDM \\
Aug   25.3810    &  0.414169  &   20.73   &  0.07 & R-band & MDM \\
Aug   25.3885    &  0.421610  &   20.59   &  0.06 & R-band & MDM \\
Aug   25.3959    &  0.429060  &   20.65   &  0.06 & R-band & MDM \\
Aug   25.4034    &  0.436510  &   20.71   &  0.07 & R-band & MDM \\
Aug   25.4108    &  0.443939  &   20.74   &  0.08 & R-band & MDM \\
Aug   25.4221    &  0.455200  &   20.57   &  0.08 & R-band & MDM \\
Aug   25.4295    &  0.462690  &   20.69   &  0.09 & R-band & MDM \\
Aug   25.4370    &  0.470129  &   20.66   &  0.09 & R-band & MDM \\
Aug   25.4444    &  0.477560  &   20.78   &  0.10 & R-band & MDM \\
Aug   25.4519    &  0.485001  &   20.83   &  0.11 & R-band & MDM \\
Aug   25.4593    &  0.492430  &   20.71   &  0.09 & R-band & MDM \\
Aug   25.4668    &  0.499910  &   20.73   &  0.11 & R-band & MDM \\
Aug   25.4742    &  0.507349  &   20.82   &  0.11 & R-band & MDM \\
Aug   25.4816    &  0.514780  &   20.98   &  0.12 & R-band & MDM \\
Aug   25.4905    &  0.523621  &   20.71   &  0.10 & R-band & MDM \\
Aug   25.9037    &  0.936850  &   21.17   &  0.09 & R-band & Maidanak \\
Aug   26.0098    &  1.04295   &   21.34   &  0.10 & R-band & NOT \\
Aug   26.0148    &  1.04795   &   21.30   &  0.06 & R-band & NOT \\
Aug   26.2076    &  1.24075   &   21.42   &  0.09 & R-band & VLT \\
Aug   26.2083    &  1.24145   &   21.45   &  0.12 & R-band & VLT \\
Aug   26.2238    &  1.25695   &   21.57   &  0.11 & R-band & D1.5m\\
Aug   26.2291    &  1.26225   &   21.47   &  0.10 & R-band & D1.5m \\
Aug   26.2349    &  1.26805   &   21.35   &  0.10 & R-band & D1.5m \\
Aug   26.2532    &  1.28635   &   21.47   &  0.07 & R-band & VLT \\
Aug   26.2921    &  1.32527   &   21.56   &  0.04 & R-band & MDM \\
Aug   26.3224    &  1.35553   &   21.67   &  0.05 & R-band & MDM \\
Aug   26.3525    &  1.38562   &   21.68   &  0.05 & R-band & MDM \\
Aug   26.3561    &  1.38925   &   21.63   &  0.15 & R-band & D1.5m \\
Aug   26.3609    &  1.39405   &   21.59   &  0.11 & R-band & D1.5m \\
Aug   26.3748    &  1.40791   &   21.71   &  0.06 & R-band & MDM \\
Aug   27.3055    &  2.33870   &   21.90   &  0.05 & R-band & MDM \\
Aug   27.3438    &  2.37690   &   21.77   &  0.05 & R-band & MDM \\
Aug   28.0352    &  3.06835   &   22.17   &  0.08 & R-band & NOT \\
Aug   28.0508    &  3.08395   &   22.09   &  0.06 & R-band & NOT \\
Aug   28.3367    &  3.36986   &   22.08   &  0.04 & R-band & MDM \\
Aug   29.3194    &  4.35259   &   22.18   &  0.05 & R-band & MDM \\
Aug   30.1455    &  5.17865   &   22.22   &  0.06 & R-band & WHT \\
Sep    2.2300    &  8.26315   &   22.64   &  0.13 & R-band & D1.5m \\
Sep   3.8139     & 9.84705    &   22.68   & 0.12 & R-band & CrAO2.6 \\
Sep   4.8392     & 10.8724    &   23.00   & 0.08 & R-band & CrAO2.6 \\
Sep   11.1326    &  17.1657   &   22.76   &  0.07 & R-band & NOT \\
Sep   14.4278    &  20.4609   &   22.95   &  0.06 & R-band & MDM \\
Sep   16.8254    &  21.8582   &   22.70   & 0.60 & R-band & Maidanak\\
Sep   29.3190    &  35.3521   &   23.59   &  0.09 & R-band & MDM \\
Sep   29.8556    &  35.8887   &   23.44   &  0.13 & R-band & NOT\\
Oct    6.1914    &  42.2245   &   23.72   &  0.09 & R-band & VLT \\
Oct  7.8490      &  43.8821   &   24.00   &  0.30 & R-band & CrAO2.6 \\
Nov  7.7356      &  74.7687   &   23.94   &  0.23 & R-band & CrAO2.6 \\
Dec    7.0694    &  104.1025   &   23.70  &  0.15 & R-band & NTT \\
\hline
Aug   25.0032   &  0.0363503  &   18.84   &  0.06  & I-band & OSN \\
Aug   25.0067   &  0.0398502  &   18.84   &  0.05  & I-band & OSN \\
Aug   25.0254   &  0.0585499  &   19.26   &  0.06  & I-band & OSN \\
Aug   25.0391   &  0.0722  &   19.11   &  0.06  & I-band & OSN \\
Aug   25.0426   &  0.0757504  &   19.38   &  0.07  & I-band & OSN \\
Aug   25.0753    &  0.108450  &   19.48   &  0.05  & I-band & NOT \\
Aug   25.0796    &  0.112749  &   19.57   &  0.07  & I-band & NOT \\
Aug   25.0870    &  0.120150  &   19.50   &  0.04  & I-band & NOT \\
Aug   25.1521    &  0.185249  &   19.84   &  0.08  & I-band & OSN \\
Aug   25.1730    &  0.206150  &   19.92   &  0.08  & I-band & OSN \\
Aug   25.1838    &  0.216949  &   19.98   &  0.09  & I-band & OSN \\
Aug   25.2602    &  0.293350  &   19.97   &  0.09  & I-band & D1.5m \\
Aug   25.2661    &  0.299250  &   19.93   &  0.08  & I-band & D1.5m \\
Aug   25.3527    &  0.385849  &   20.10   &  0.07  & I-band & D1.5m \\
Aug   25.3659    &  0.399050  &   20.12   &  0.08  & I-band & D1.5m \\
\hline
Aug   26.3472    &  1.3804    &   19.03   &  0.05  & K-band & VLT \\
\end{longtable}

\noindent
\clearpage


\begin{table}
\caption{Local calibration stars.
\label{t:stars}}
\centering   
\begin{tabular}{l c c c c}
\hline\hline  
ID & $B$ & $V$ & $R$ & $I$ \\
\hline
Star A & 17.42 & 16.60 & 16.18 & 15.60 \\
Star B & 18.17 & 17.29 & 16.83 & 16.18 \\
Star C & -- & 20.23 & 19.14 & 17.75 \\
Star D & -- & 20.66 & 19.54 & 17.77 \\
\hline
\end{tabular}
\end{table}

\begin{table}
\caption{Spectral line measurements.
\label{t:lines}}
\centering   
\begin{tabular}{l c c c c}
\hline\hline  
ID & Rest Wavelength & Observed Wavelength & Flux & Redshift  \\    
   &  (\AA) &           (\AA) & (10$^{-17}$ erg s$^{-1}$ cm$^{-2})$ & \\    
\hline
[\ion{O}{II}] &  3727.42    &  6812.94    &   3.5    & 0.828 \\

[\ion{Ne}{III}] &    3868.75  &    7075.00  &     2.0  &   0.828 \\

H$\beta$ &     4861.33 &     8885.96 &    2.4 &    0.828 \\

[\ion{O}{III}]  &   4958.91   &   9064.99   &    6.0   &  0.828 \\

[\ion{O}{III}]   &   5006.84   &   9152.43   &    14.7   &  0.828 \\ 

Mg II  &    2800.0  &    5117.63  &     --    &   -- \\
\hline
\end{tabular}
\end{table}


\end{document}